# Assisting V2V failure recovery using Device-to-Device Communications

Emad Abd-Elrahman*, Adel Mounir Said*, Thouraya Toukabri **, Hossam Afifi*** and Michel Marot***
*National Telecommunication Institute, Cairo, Egypt; ** Orange Labs, Issy Les Moulineaux, France; *** Institut Mines-Telecom, RST Department, Saclay, France.
{emad.abd_elrahman, hossam.afifi, michel.marot}@telecom-sudparis.eu, amounir@nti.sci.eg, thouraya.toukabrigunes@orange.com

*Abstract*— This paper aims to propose a new solution for failure recovery (*dead-ends*) in Vehicle to Vehicle (V2V) communications through LTE-assisted Device-to-Device communications (D2D). Based on the enhanced networking capabilities offered by Intelligent Transportation Systems (ITS) architecture, our solution can efficiently assist V2V communications in failure recovery situations. We also derive an analytical model to evaluate generic V2V routing recovery failures. Moreover, the proposed hybrid model is simulated and compared to the generic model under different constrains of worst and best cases of D2D discovery and communication. According to our comparison and simulation results, the hybrid model decreases the delay for alarm message propagation to the destination (typically the Traffic Control Center TCC) through the Road Side Unit (RSU).

*Keywords—Vehicular Networks VANET; ITS; Modeling; D2D Communication*

## I. THE HYBRID ARCHITECTURE

Pure vehicular communications also known as V2V can suffer from blocking or failure: '*dead-ends*' [1]. This is mainly due to geographical guidance errors, communication range being too short, or dynamicity in the topology.

Our work is inspired from the ITS architecture principles [2] and the new LTE-based D2D communication mechanisms and offers a new cognitive model that solves V2V blocking and failure (the D2D mechanism used in our hybrid model is inspired from [3]). The idea is to take profit from the extended ITS network management features allowing vertical handovers between different access media to recover '*dead-ends*' V2V failures as shown in Fig. 1. We believe that a mixed architecture combining ad-hoc V2V and intermediate D2D communication improves the overall transmission success ratio and delay. The D2D support can be seen as a failover solution that could be a little bit slower than direct V2V in the worst case (*discovery phase is done on demand*), or which could help interconnecting disconnected groups of mobile nodes and enhance the processing delay in the best case (*discovery phase is done proactively*).

In VANET, some contributions study the messages relaying and spraying copies over networks while others study the routing issue from a geographical aspect using geo-location and GPS [4]. The problem of '*dead ends*' in VANET happens when the routing disconnects due to a '*hole to next hop*'. In this case a compensation mechanism can be used to help V2V packet routing along alternative paths. Proposed ways depend on either go-back '*one*' or '*many*' hops to find another relay for messages. Other solutions tried to overcome this problem by redirecting the packet in the reverse way in order to find an alternative path to RSU. Many ad-hoc protocols deal with the '*dead ends*' problem and are explained in [5]. Besides, parallel paths are considered as backup solutions for V2V routing failure [1].

In addition to solving the '*hole to next hop*' problem, this contribution details a new method to analytically model generic V2V communications. The models that are crosschecked with simulations confirm the robustness of our hybrid protocol and the strength of using models instead (or in addition) of simulations. In the following, the presentation order of these two ideas is inversed because we want to follow a logical scenario highlighting the methodology before explaining the solution.

A *generic* routing approach is used to compare our hybrid algorithm based on D2D with many other techniques in a generic way based only on important and significant criteria. Our goal is to prove that our D2D recovery process in V2V failure cases is more efficient than traditional V2Vtechniques. We concentrate our comparison on geographical routing to find an alternative path in case of failure.

In a failure situation, we assume that the vehicle (*x*) at a given position and time ($T_f$) has no neighbor to relay the alert message it received from its predecessor using any routing algorithm. Then, a D2D session will be established thanks to the eNodeB to compensate this failure as shown in Fig. 1.

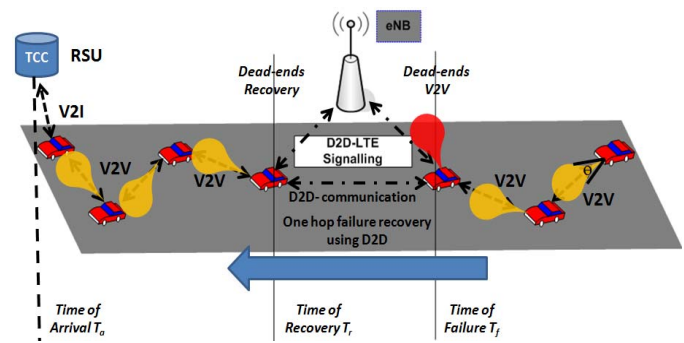

Fig. 1. Basic hybrid model of V2V assisted by D2D.

## II. ANALYTICAL MODEL

We evaluate the performance of our hybrid solution with two distinct approaches and we compare both methods together at the end. The first approach is analytical. We believe that with an analytical model, even if approximated, results can be much more beneficial and can be easily integrated in real-time on board systems for real implementation and usage purposes. This generic V2X technique could be applied with slight modifications for any V2V or V2I communication algorithm. The second approach is via simulation. It gives more extensive results and can be adapted to any scenario, so this is considered as a longer term result analysis method.

### A. Analytical V2V Connected Set Model

We consider a free flow vehicles' traffic on a straight road with speeds distributed according to a truncated normal distribution. The minimum and maximum speeds are $V_{min}$ and $V_{max}$, their mean and standard deviation are $\mu$ and $\sigma$, the transmission range $R$. The vehicles arrive at position 0 of the road according to a Poisson process with rate $\lambda_a$. As detailed in [6], the inter-distance of the vehicles at any given time is exponentially distributed with rate as follows:

$$\lambda = \lambda_a \int_{V_{min}}^{V_{max}} \frac{1}{v} \frac{\frac{2}{\sigma\sqrt{2\pi}} e^{-\frac{1}{2}(\frac{v-\mu}{\sigma})^2}}{\text{erf}(\frac{V_{max}-\mu}{\sigma\sqrt{2}}) - \text{erf}(\frac{V_{min}-\mu}{\sigma\sqrt{2}})} dv \quad (1)$$

We adopt the same approach for modeling the connectivity in terms of the GI$^x$/D/∞ as detailed in [6]. As we consider the number of hops on the shortest path, we take into account the retransmitting nodes only instead of all the vehicles on the road. Thus, the input process of the queue is no more the Poisson process but a sampled process on this Poisson process, which is a Markovian process.

To calculate the inter-distance between the retransmitting vehicles on the road, we assume a shortest path between them as in Fig. 2. $X_1, X_2,...,X_n$. $X_{n+1}$ is the farthest vehicle under $X_n$'s coverage, or the next node after $X_n$ if the distance between $X_n$ and $X_{n+1}$ is larger than $R$. If $X_{n+1}$ is the farthest vehicle under $X_n$ coverage, there may be other vehicles between $X_n$ and $X_{n+1}$ but we assume they do not retransmit the packets sent by $X_n$. The $X_{n+1}$ is assumed to be the next hop for the $X_n$'s as it is within its coverage range. Let $\tau_n$ be the distance between node $X_n$ and node $X_{n+1}$. Let $N[a;b]$ be the number of vehicles between positions a and b when $\forall x_n \leq R$, the probability function of $\tau_n$ knowing $\tau_{n-1}$ is given by (replace $\tau_0 = R$ to get ($\tau_1 \leq x_1$)):

$$F_{\tau_n/\tau_{n-1}}(x_n, x_{n-1}) = P(\tau_n \leq x_n / \tau_{n-1} = x_{n-1} \cap \tau_{n-1} \leq R)$$
$$= e^{-\lambda(R-x_n)} - e^{-\lambda x_{n-1}} \quad (2)$$

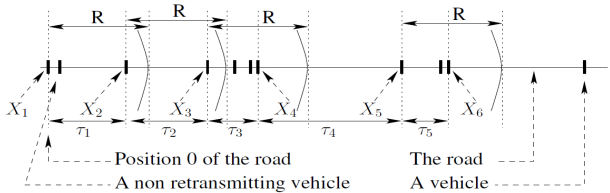

Fig. 2. Vehicle connected component according to the shortest path.

The work of a GI$^x$/D/∞ considered queue whose customers arrive in groups with a generating function $A(z)$ separated by inter-arrival times $\tau_n$ with the distribution function $F(x)$ and the service time distribution is $H(x)$. $Y_n$ is the number of clients in the $n^{th}$ group and $T_n$ is the time necessary to complete the services of that group. $K(x) = A(H(x))$ is the distribution function of $T_n$. They show that the z-transform of the distribution function of the number $N_b$ of clients in a connected component is:

$$\sum_{k=1}^{+\infty} P(N_b = k) z^k =$$
$$\sum_{n=1}^{\infty} [A(z)]^n \int_0^\infty \cdots \int_0^\infty \prod_{i=1}^{n-1}\{[K(\sum_{j=i}^n x_i) - K(x_i)]dF(x_i)\} K(x_n) dF(x_n) \quad (3)$$

Considering the special case of the GI$^x$/D/∞, when

$$K(x) = \begin{cases} 1 & \text{if } x \geq R \\ 0 & \text{otherwise} \end{cases} \quad (4)$$

Equation (3) will be simplified to:

$$\sum_{k=1}^{+\infty} P(N_b = k) z^k = z \frac{1-F(R)}{1-zF(R)} \quad (5)$$

As the inter-arrivals of the GI$^x$/D/∞ are independent, the formula of (3) can be simplified as in (5). But, the inter-distance between the vehicles is not an independent process as we assumed to use the shortest path between the transmitting vehicles, which leads to a Markovian process as mentioned before. Consequently, the computing of the probability of the number of vehicles in a connected component cannot be calculated depending on the results that rely on the GI$^x$/D/∞ queue. Therefore, the results must be adapted to a non-independence case.

Using theorem 1 in [6], the probability to have k vehicles in a connected component is:

$$P(N_b = k) = \int_{x_1=0}^{R} \int_{x_2=R-x_1}^{R} \cdots \int_{x_{k-1}=R-x_{k-2}}^{R} P(\tau_k > R / \tau_{k-1} = x_{k-1}) \times$$
$$\prod_{i=2}^{k-1} dP(\tau_i = x_i/\tau_{i-1} = x_{i-1}) dP(\tau_1 = x_1) \quad (6)$$

Let us denote $\lambda' = \lambda R$, $\rho = \lambda' e^{-\lambda'}$, and $\rho' = e^{-\lambda'}$.

With a simple change of variable, (6) simplifies into

$$P(N_b = k) = \rho^{k-1} \times$$
$$\int_{u_1=0}^{1} \int_{u_2=1-u_1}^{1} \cdots \int_{u_{k-1}=1-u_{k-2}}^{1} \prod_{i=1}^{k-2} e^{-\lambda' u_i} \prod_{i=1}^{k-1} du_i \quad (7)$$

By denoting

$$\mathfrak{M}_{\alpha,k} =$$
$$\rho^k \int_{u_1=0}^{1} \int_{u_2=1-u_1}^{1} \cdots \int_{u_k=1-u_{k-1}}^{1} \prod_{i=1}^{k-1} e^{-\lambda' u_i} \prod_{i=1}^{k} du_i \quad (8)$$

We have: $P(N_b = k) = \rho \mathfrak{M}_{1,k-2} \quad (9)$

Then, by defining the following z-transforms,:

$$Q(z) = \sum_{k=1}^{+\infty} P(N_b = k) z^k \quad (10)$$

$$M_1(z) = \sum_{k=1}^{+\infty} \mathfrak{M}_{1,k} z^k \quad (11)$$

Obviously, the z-transform of $N_b$ is:

$$Q(z) = \rho' + \rho z^2(1 + M_1(z)) \quad (12)$$

And assuming a special case where $\lambda R \geq Ln4$, it can be proved (cf. [6]) that the z-transform $M_1(z)$ can be expressed as:

$$M_1(z) = \frac{h_1(z)+h_2(z)+h_3(z)}{\rho z^2\left[1+\sqrt{1-4\rho' z^2}-2z e^{\frac{1}{2}\lambda'\left(\sqrt{1-4\rho' z^2}-1\right)}\right]} \quad (13)$$

where

$$h_1(z) = \sqrt{1-4\rho' z^2}[(1-\rho'-\rho)z^3 - (1-\rho')z^2 \\ -z(1-\rho) + 2 - \rho' - \rho] \quad (14)$$

$$h_2(z) = e^{\frac{1}{2}\lambda'\left(\sqrt{1-4\rho' z^2}-1\right)}[2\rho z^3 + 2\rho' z^2 - z - 1 \\ +(z-1)\sqrt{1-4\rho' z^2}] \quad (15)$$

$$h_3(z) = z^3(\rho'+\rho-1) + z^2(1-3\rho'-2\rho) \\ +z(1-\rho) + \rho' + \rho \quad (16)$$

The number of hops in the connected component for an infinite road length (i.e. without board effects) can be calculated by differentiating (13) with respect to z as:

$$E(no. of\ hops) = \sum_{k=1}^{+\infty} kP(N_b = k)$$
$$= P(N_b = 1) + 2P(N_b = 2) + \rho\left(\frac{\partial M_1(z)}{\partial z}\right)_{z=1}$$
$$= \rho' + 2\rho + 2\rho M_1(1) + \rho\left(\frac{\partial M_1(z)}{\partial z}\right)_{z=1} \quad (17)$$

The average size of the connected component can be calculated as (cf. [6] for details in references):

$$E(size) = \frac{1}{\lambda e^{-\lambda R}} - \frac{1}{\lambda} \quad (18)$$

Finally, we can calculate the number of hops over a road of length L:

$$E(no. of\ hops)_L = \frac{E(no.of\ hops)}{E(size)} \times L \quad (19)$$

Note that having the z-transform of $N_b$ allows obtaining any moment for accurate results of performance evaluation.

### B. Preliminary Evaluation via Simulation

Fig. 3 shows our results for different communication ranges based on our analytical model versus simulation using the same model and parameters. As shown in the figure, the delay increases while the number of hops is almost constant. This is due to the increase of the vehicle density, and hence the access delay increases. It is clear that both results are very close in terms of number of hops and approximate delay.

Fig. 4 shows preliminary simulations for different D2D use cases; it seems at the first glance that V2V gives a shorter delay. However, the overall end-to-end delay in D2D cases is better than V2V delay regardless to the backward hops number as LTE-based D2D communication range is always 3 to 5 times larger than V2V range (IEEE 802.11 family).

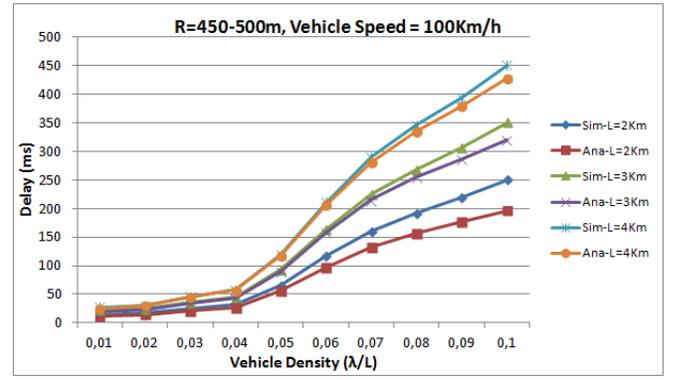

Fig. 3. Analytical versus simulation results for the number of hops & delay based on different road lenghts and communication range.

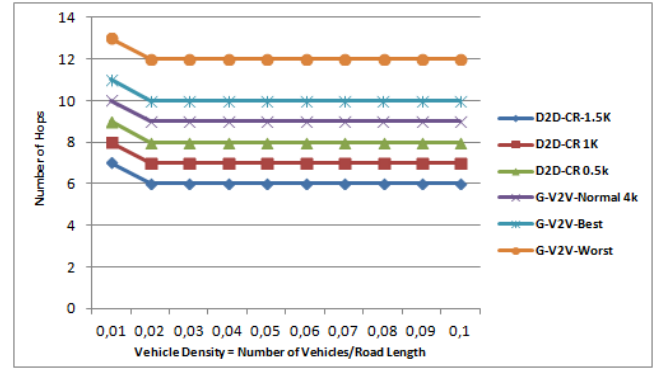

Fig. 4. Simulation results for D2D use cases with different communication range (CR) versus pure V2V routing recovery in terms of number of hops.

### III. CONCLUSIONS

This work proposed a complete framework to compare a new ITS paradigm for V2V communication failure system using LTE-based D2D communications. First, we introduced a new complete analytical model for V2V continuity messages. Moreover, the analytical results for the generic routing algorithm have been analyzed in terms of hops number and delay. Then, a hybrid model based on D2D-assisted V2V is explained and simulated. We show an improvement in delay in the overall path for messages propagation by D2D use case.